\documentclass[twocolumn,amsmath,amssymb,longbibliography,superscriptaddress,aps,pra,10pt]{revtex4-1}

\usepackage{graphicx}

\usepackage[english]{babel}

\usepackage[T1]{fontenc}
\usepackage[utf8]{inputenc} 
\usepackage{lmodern}
\usepackage{microtype}

\usepackage{chemformula}
\usepackage{siunitx}
\usepackage{bm}

\usepackage{color}
\usepackage[colorlinks=true,allcolors=blue]{hyperref}

\newcommand{\Ar}[1]{$^{#1}$Ar}
\newcommand{\Kr}[1]{$^{#1}$Kr}
\newcommand{\C}[1]{$^{#1}$C}

\begin{document}
\renewcommand{\figurename}{\textbf{Figure}}

\title{\texorpdfstring{\Ar{39} dating with small samples resolves ocean ventilation}{Ar-39 dating with small samples resolves ocean ventilation}}

\date{\today}

\author{S. Ebser}
\email{oceanArTTA@matterwave.de}
\affiliation{Kirchhoff-Institute for Physics, Heidelberg University, 69120 Heidelberg, Germany}

\author{A. Kersting}
\affiliation{Institute of Environmental Physics, Heidelberg University, 69120 Heidelberg, Germany}

\author{T. Stöven} 
\affiliation{GEOMAR Helmholtz-Centre for Ocean Research Kiel, 24148 Kiel, Germany}

\author{Z. Feng} 
\author{L. Ringena} 
\affiliation{Kirchhoff-Institute for Physics, Heidelberg University, 69120 Heidelberg, Germany}

\author{M. Schmidt}
\affiliation{Kirchhoff-Institute for Physics, Heidelberg University, 69120 Heidelberg, Germany}
\affiliation{Institute of Environmental Physics, Heidelberg University, 69120 Heidelberg, Germany}

\author{T. Tanhua}
\affiliation{GEOMAR Helmholtz-Centre for Ocean Research Kiel, 24148 Kiel, Germany}

\author{W. Aeschbach}
\affiliation{Institute of Environmental Physics, Heidelberg University, 69120 Heidelberg, Germany}
\affiliation{Heidelberg Center for the Environment HCE, Heidelberg University, 69120 Heidelberg, Germany}

\author{M. K. Oberthaler}
\affiliation{Kirchhoff-Institute for Physics, Heidelberg University, 69120 Heidelberg, Germany}

\pacs{}

\begin{abstract}
Ocean ventilation is the integrated effect of various processes that propagate surface properties to the ocean interior. Its precise understanding is the prerequisite for addressing essential questions such as oxygen supply, the storage of anthropogenic carbon and the heat budget of the ocean. Currently employed observational methods to infer ventilation utilise transient tracers, i.e. tritium, \ch{SF6}, CFCs and the radioisotope \C{14}. However, their dating ranges are not suitable to resolve the dynamics of the deep ocean. The noble gas isotope \Ar{39} with a half-life of 269 years fills this gap. Its broad application has previously been hindered by its very low abundance, requiring at least 1000 litres of water for dating. Here we report on successful \Ar{39} dating with only 5 litres of water based on the atom-optical technique Atom Trap Trace Analysis. Our data reveal previously not quantifiable ventilation patterns in the Eastern Tropical North Atlantic, where we find that advection is more important for the ventilation of the intermediate depth range than previously assumed. This result implies faster ventilation than estimated in other studies and thus a significantly higher anthropogenic \ch{CO2}-storage. The demonstrated analytical capabilities now allow for a global collection of \Ar{39} data, which will have significant impact on our understanding of ocean ventilation. 
\end{abstract}

\maketitle

The well mixed surface layer of the ocean exchanges properties with the atmosphere through air-sea gas exchange. Various processes such as advection and diffusion are responsible for transporting surface waters with their corresponding properties to the ocean interior. The integrated effect of such water mass propagation is termed ocean ventilation. Knowledge of its temporal and spatial variations is essential for a reliable prediction of the Earth system’s response to climate change\cite{IPCC, Broecker1982}. Thus, a systematic observation on a global scale is desired. With the new capabilities reported here, this now appears to be feasible.

Quantifying ventilation implies an estimation of the time since the water was last in contact with the atmosphere, i.e. the age of the water. It can be accessed by observations of transient tracers which encode time information via radioactive decay or a time dependent input function\cite{Stoeven2015}. The short atmospheric histories of well-established transient tracers such as chlorofluorocarbons (CFCs) and sulfurhexafluoride (\ch{SF6})\cite{Fine2011} cover only the last 70 years, preventing the dating of the slowly ventilated part of the global ocean. There, time-scales of approximately 1000 years are estimated from \C{14} measurements\cite{Matsumoto2007}. However, these results are rather uncertain due to the complex carbon dynamics, long air-sea equilibration time and long half-life compared to ocean ventilation. The chemically inert noble gas isotope \Ar{39} with a half-life of 269 years has long been identified as the ideal tracer for the time scales of deep ocean circulation\cite{Loosli1983, Loosli1968, Broecker2000, Holzer2010, Schlosser1994}. However, even though argon is a common gas in the atmosphere, the desired \Ar{39} isotope is extremely rare, due to its very low isotopic abundance of \num{8e-16}. The corresponding low activity necessitates samples of \num{\sim1000} litres of water for \Ar{39} detection by low-level counting (LLC) of the radioactive decays\cite{Loosli1983, Schlosser1994, Schlitzer1985}. This large sample size hinders routine measurements of ocean samples. Here we show that our method for analysing \Ar{39}, which we call Argon Trap Trace Analysis (ArTTA), reduces the required water volume to 5 litres. This makes large scale ocean surveys feasible as taking 5 litres of water can be readily integrated into standard water sampling procedures on research vessels.

\begin{figure*}[htb]
\includegraphics[width=1\textwidth]{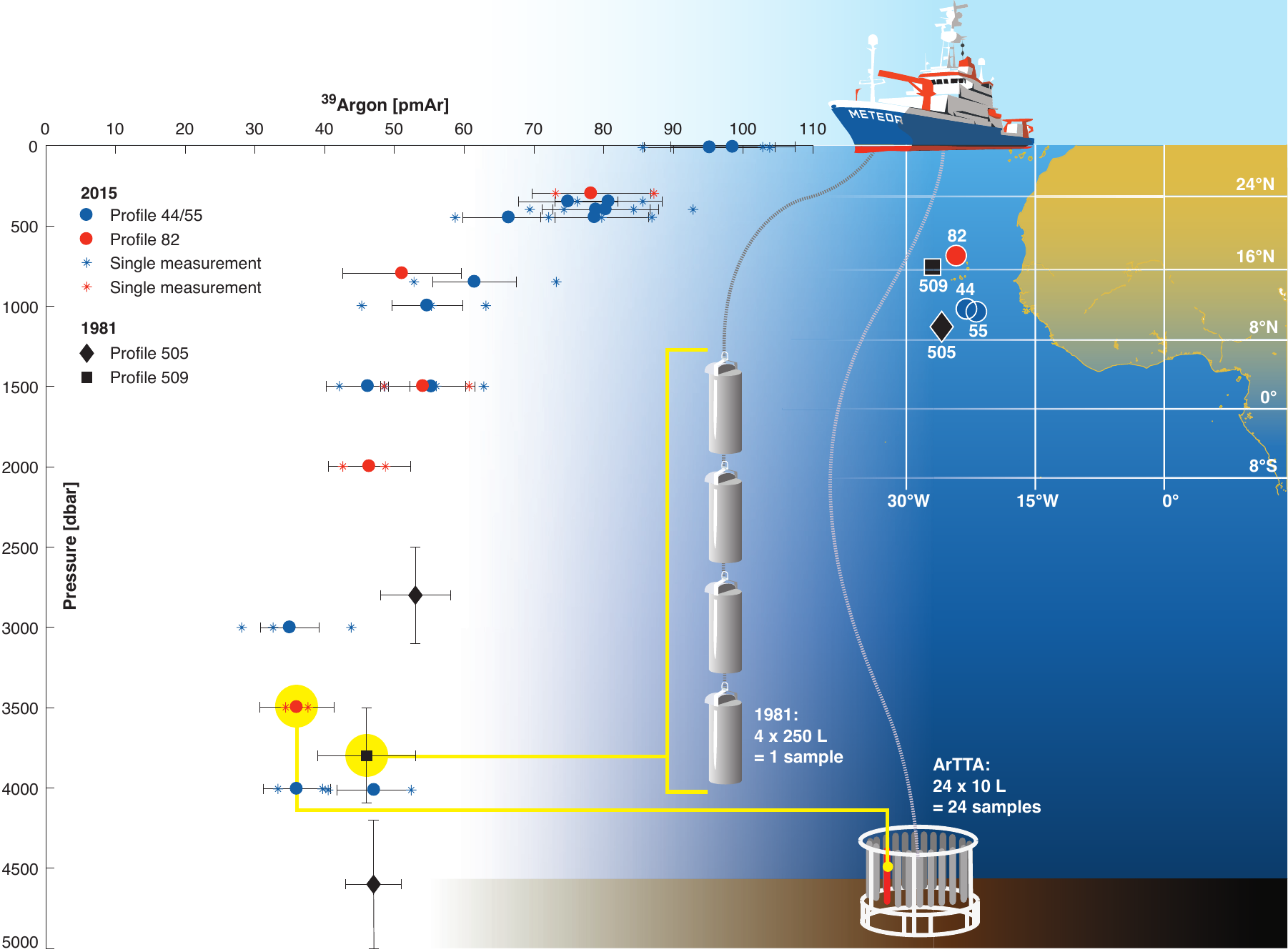}
\caption{\textbf{\Ar{39} concentrations in percent of modern Argon (pmAr) for different depths in the Eastern Tropical North Atlantic.} Each sample was analysed with at least two independent measurements and the horizontal error bars represent the uncertainties of the combined single measurements. The sensitivity and efficiency of ArTTA allows for a full depth profile from just one CTD-cast corresponding to sample sizes as small as 5 litres of water. For comparison, three historic (1981) data points\cite{Schlitzer1985} from two specific sites are included; the required 1000 litres for LLC analysis were obtained by combining four 250 litres sampling bottles integrating over a depth range of 600 to 800 meters (vertical error bars).}
\label{fig:fig1}
\end{figure*}

\section*{Argon Trap Trace Analysis}
For the detection of long-lived radioisotopes, it is more efficient to count atoms rather than radioactive decays\cite{Collon2004}. For example, atom counting by accelerator mass spectrometry (AMS) dramatically reduced sample-size requirements for \C{14} dating. AMS is not easily applicable for noble gases, yet atom counting and the related substantial reduction of sample size becomes possible by employing the modern atom-optical technique known as Atom Trap Trace Analysis (ATTA). It utilises techniques from the field of atomic physics to detect rare isotopes down to the \num{e-16}-level. It exploits shifts of the optical resonance frequency due to different isotopic mass and nuclear spin. The high background of abundant isotopes hinders selection by a single resonant excitation, but the sensitivity is strongly enhanced by many cycles of photon absorption and subsequent spontaneous emission. Therefore, ATTA’s outstanding isotopic selectivity is based on millions of resonant photon scattering events required for cooling, trapping and detecting single atoms inside a magneto-optical trap (MOT). 

The general concept of ATTA has first been demonstrated for the rare isotopes \Kr{85} (half-life of \num{10.76} years) and \Kr{81} (half-life of \num{229000} years)\cite{Chen1999, Jiang2012} and is applied for dating groundwater\cite{Sturchio2004, Aggarwal2014, Gerber2017} and ice\cite{Buizert2014}. While the first \Ar{39} detection by this approach was reported in a proof of concept experiment in 2011 \cite{Jiang2011}, the first explicit demonstration for dating groundwater samples was achieved in 2014 \cite{Ritterbusch2014}. The first apparatus used in that study still required \num{\sim 1000} litres of water. Since then the setup has been significantly improved by doubling the count rate, ensuring reliability and employing well characterised enriched reference samples. The crucial step for the reduction of the necessary sample size is the implementation of gas recirculation in an optimised vacuum system. With this ArTTA system, \Ar{39} quantification is now possible with only \SI{2}{mL\;STP} of argon, which can be extracted from 5 litres of water. Thus, a complete ocean depth profile of \Ar{39} can be sampled from one standard hydrographic cast equipped with \SI{10}{\L} Niskin sampling bottles, fulfilling the requirements for broader application of \Ar{39} in oceanography\cite{Lu2014}.

\begin{figure*}[htb]
\includegraphics[width=1\textwidth]{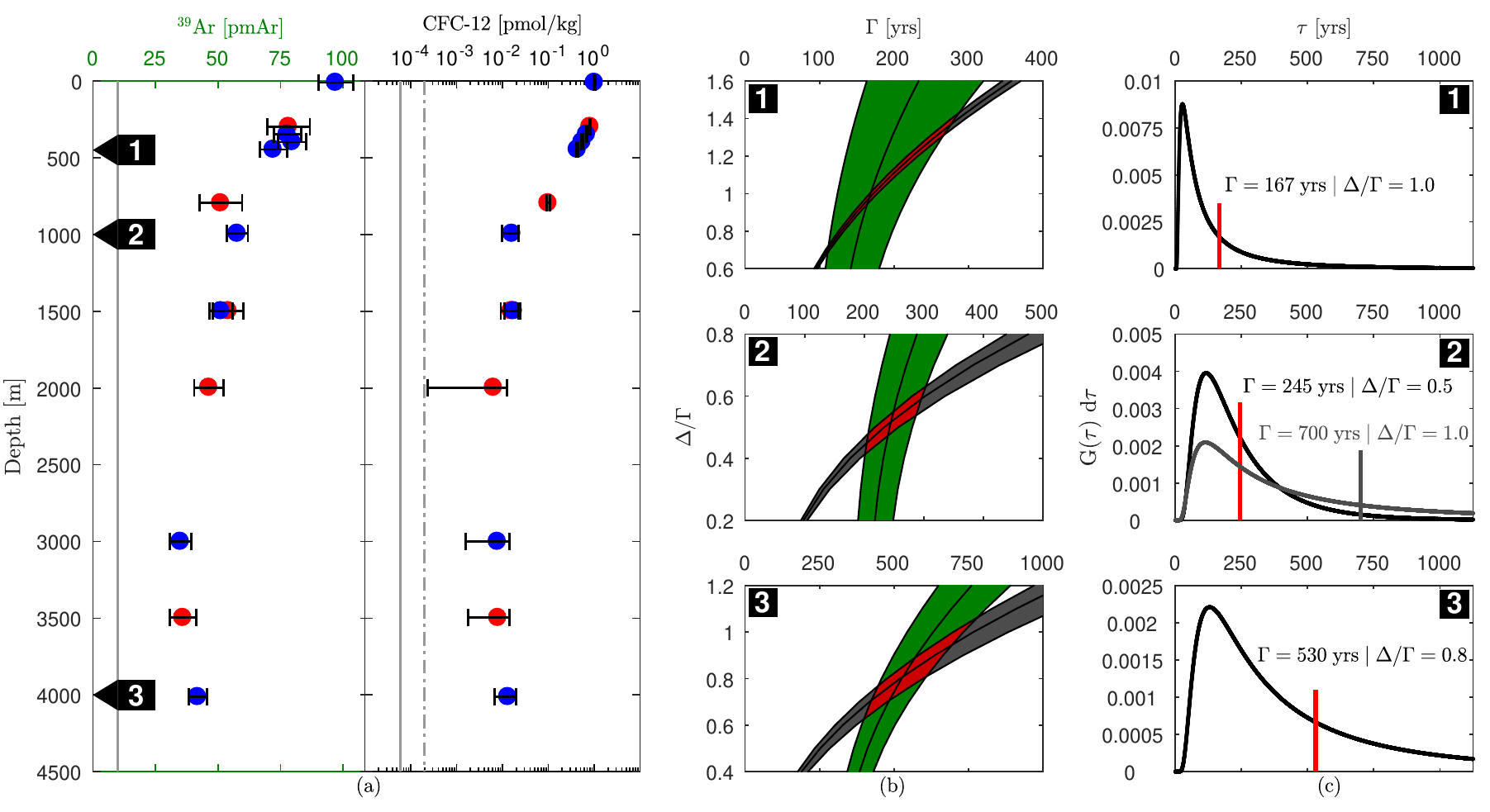}
\caption{\textbf{Constraining Inverse Gaussian Transit Time Distributions (IG-TTD).} \textbf{Panel a)} shows the concentration profiles of \Ar{39} on a linear and of CFC-12 on a logarithmic scale: Red markers for profile \#82, blue markers for the combined profile \#44/\#55. The indicated uncertainties for \Ar{39} are given by the counting statistics, and for CFC-12 due to the on-board system performance and the known systematic uncertainties. The solid and dashed grey vertical lines indicate the limits of detection and quantification, respectively. \textbf{Panel b)} illustrates how \Ar{39} (green) and CFC-12 (grey) data constrain possible parameter combinations $\left(\Gamma, \nicefrac{\Delta}{\Gamma}\right)$ of the IG-TTD given by their intersecting red area. \textbf{Panel c)} shows the corresponding distributions. For comparison we add in the middle panel the distribution obtained using only CFC-12 data with the commonly assumed $\nicefrac{\Delta}{\Gamma}$ unity ratio.}
\label{fig:fig2}
\end{figure*}

\section*{\texorpdfstring{\Ar{39} depth profiles}{Ar-39 depth profiles}}
Here we apply our new analytical capabilities to explore the ventilation regime in the Eastern Tropical North Atlantic, in the context of investigations of the Oxygen Minimum Zone in this region. In 2015, three depth profiles were taken with only one single hydrographic cast per profile during the research cruise M116 on research vessel Meteor. Two profiles (\#44, \#55) originate from the centre of the Oxygen Minimum Zone and one profile (\#82) from the Cape Verde Ocean Observatory\cite{WebpageGeomar} (Figure \ref{fig:fig1}). The ocean water was sampled by closing three 10 litres Niskin bottles of a 24 bottle rosette per depth and transferring \num{\sim 7} litres of the content of each Niskin bottle into an evacuated 27 litres commercial propane gas bottle. This way, 24 sampling containers filled with 20 litres of water from eight different depths at three sampling sites were taken. Gas extraction from these samples followed by argon separation resulted in a total argon yield between \num{5} and \SI{8}{\mL\;STP}, consistent with the temperature dependent solubility\cite{Weiss1970}. Each sample was analysed by at least two independent measurements, where the analytical uncertainties are dominated by counting statistics. As seen in Figure \ref{fig:fig1}, we find a clear decline in \Ar{39}-concentration in the upper 1000 meters, a minimum at \num{\sim 3000} meters and a slight increase towards the bottom at 4000 meters. For comparison, we also include the three historic \Ar{39} samples closest to our sampling positions, which were taken in 1981 and analysed by low-level counting\cite{Schlitzer1985}. The observed differences in concentrations are consistent with the known meridional gradients in tracer concentrations characteristic for this region. Note that these historic \Ar{39} data are integrated concentrations of a depth interval of 600 to 800 meters as indicated by the vertical error bars due to the required large sample size of 1000 litres. The ArTTA technique allows for 24 \Ar{39} samples during one single cast (with a “standard” 24 bottle rosette) that takes \num{\sim 3} hours of ship-time, which is about the same amount of time needed for one large volume sample required for LLC. 

\section*{Transit Time Distribution}
In the following we discuss how the obtained \Ar{39} concentrations constrain the ventilation in this area. Since ventilation always implies mixing along the flow path, it is adequate to describe a water sample by a distribution of ages rather than one distinct age. One common assumption is that this age distribution, known as Transit Time Distribution (TTD)\cite{Khatiwala2001, Hall1994}, is described by the Inverse Gaussian function (IG-TTD)\cite{Waugh2003}, which is defined by its first two moments: The mean of the distribution $\Gamma$ and the width of the distribution $\Delta$

$$G\left(\tau\right) = \sqrt{\frac{\Gamma^3}{4\pi\Delta^2 \Gamma^3}} \exp{\frac{-\Gamma\left(\tau-\Gamma\right)^2}{4\Delta^2 \tau}}.$$

This TTD is adequate for oceanic and atmospheric transport described by a one dimensional flow model with constant advective velocity and diffusivity. The $\nicefrac{\Delta}{\Gamma}$ ratio indicates the diffusive compared to the advective transport characteristics of a water parcel; the lower $\nicefrac{\Delta}{\Gamma}$ the more dominant the advection. A $\nicefrac{\Delta}{\Gamma}$ ratio of 0.4-0.8 indicates an advectively dominated transport, whereas a high ratio of 1.2-1.8 indicates transport dominated by diffusive processes. Several approaches have been proposed for constraining the parameters of the IG-TTD based on tracer data; here we use the method outlined and thoroughly discussed by\cite{Stoeven2014}. This approach is based on constraining the TTD for each sample individually. Given a TTD for a location in the ocean interior, the concentration at sampling time $t_s$ and location $\bm{r}$ is described by

$$ c\left(t_s,\bm{r}\right)  = \int_0^\infty \! c_0\left(t_s-\tau\right)\cdot e^{-\lambda\tau}\cdot G\left(\tau,\bm{r}\right)\, \mathrm{d}\tau $$

where $c_0 \left(t_s-\tau\right)$ corresponds to the concentration at source year $t_s - \tau$ related to the input function of a tracer. The exponential term accounts for the decay rate of radioactive transient tracers. $G(\tau,\bm{r})$ is the Green’s function of the flow model for the given location\cite{Hall1994}, in our case the IG-TTD. The two parameters of the IG-TTD can be constrained using observed concentrations of at least two transient tracers with sufficiently different time information covering the expected age range of the water parcel\cite{Stoeven2015, Waugh2003}.

As second independent tracer we choose CFC-12, which was sampled at the same, or nearby, position. In Figure \ref{fig:fig2} a) the corresponding \Ar{39} and CFC-12 concentrations are depicted and are well above the limit of quantification. The uncertainties of the \Ar{39} concentrations are limited by the counting statistics and for CFC-12 by the system performance of the on-board measurements and the known systematic uncertainties. In Figure \ref{fig:fig2} b) we illustrate for three different depths how both measured tracers constrain possible parameter combinations $\left(\Gamma, \nicefrac{\Delta}{\Gamma}\right)$. Considering the year of sampling and the input function of the tracers, a particular tracer concentration can correspond to a range of combinations of $\Delta$ and $\Gamma$, which we plot as $\nicefrac{\Delta}{\Gamma}$ vs. $\Gamma$ \cite{Stoeven2014}. For two (or more) independent tracers multiple such curves can be analysed and the range of possible combinations of $\Delta$ and $\Gamma$ can be constrained, as indicated by the intersecting areas. The precision of the results is limited by the combination of the uncertainty of the input function (including saturation of the tracer at water mass formation), and the analytical and interpolation uncertainties. For our study we assume for \Ar{39} a constant input and for CFC-12 reported data for the northern hemisphere\cite{Bullister2017}. In Figure \ref{fig:fig2} c) the corresponding IG-TTDs are depicted.
 
 \begin{figure*}[htb]
\includegraphics[width=1\textwidth]{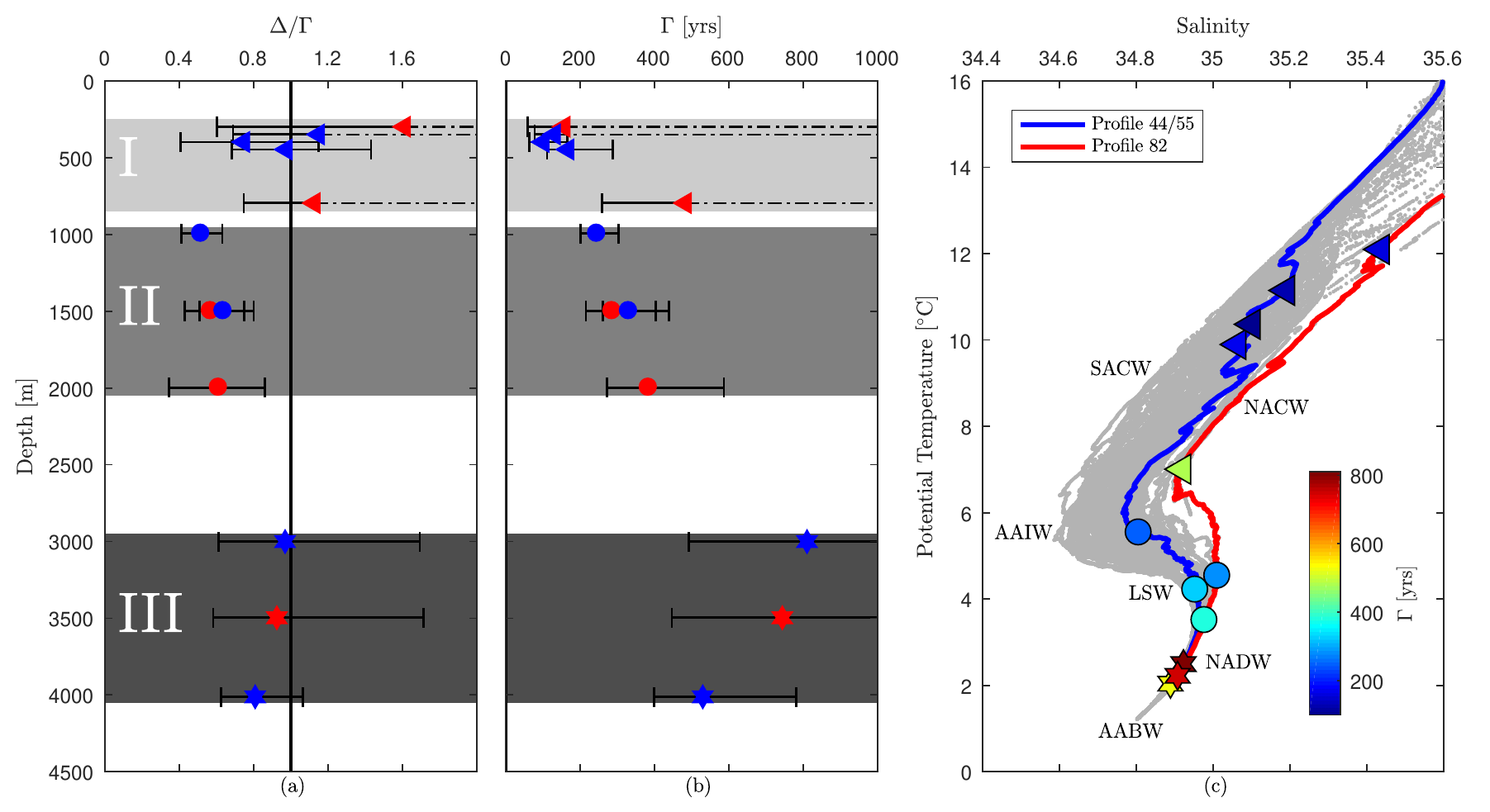}
\caption{\textbf{Identified ventilation regimes for the Eastern Tropical North Atlantic.} Estimated $\left(\nicefrac{\Delta}{\Gamma}\right)$ ratios \textbf{(a)} and mean ages $\Gamma$ \textbf{(b)}. We identify three different ventilation regimes as indicated, characterised by $\left(\nicefrac{\Delta}{\Gamma}\right) \approx 1$ above 800 meters and below 3000 meters and $\left(\nicefrac{\Delta}{\Gamma}\right) \approx 0.5$ in the intermediate regime. \textbf{c)} Temperature versus salinity of the study area: The grey dots represent the temperature and salinity data of the whole cruise, while the blue and red lines highlight the S/T curves for profiles \#44/\#55 and \#82, respectively. AABW – Antarctic Bottom Water, NADW – North Atlantic Deep Water, LSW – Labrador Sea Water, AAIW – Antarctic Intermediate Water, SACW – South Atlantic Central Water, NACW – North Atlantic Central Water.}
\label{fig:fig3}
\end{figure*}
 
\section*{Implications for ventilation regimes}
The water column can be tied to different regimes based on ventilation as described by $\nicefrac{\Delta}{\Gamma}$ ratios and mean ages $\Gamma$ (Figure \ref{fig:fig3} a and b) or by the dominant water masses as illustrated by their salinity and temperature characteristics (Figure \ref{fig:fig3} c). The water samples above \num{\sim800} meters correspond to the Atlantic Central Waters. Here we find a unity ratio $\nicefrac{\Delta}{\Gamma} = 1 $ and mean ages consistent to estimates based on the tracer couple \ch{SF6}/CFC-12 for this ocean region\cite{Schneider2012}. It is important to note that this regime is only poorly constrained by \Ar{39}. For deeper and thus older water the \ch{SF6}/CFC-12 tracer couple is not applicable for determining the age distribution, but \Ar{39}/CFC-12 gives new insights. For example, in intermediate depths (1000-2000 meters), where Antarctic Intermediate Water (AAIW) and Labrador Sea Water (LSW) dominate, we find $\nicefrac{\Delta}{\Gamma} =  0.5 - 0.6$ based on our data. This is in stark contrast to the $\nicefrac{\Delta}{\Gamma}$ unity ratio, which has been commonly applied to this depth interval before\cite{Schneider2012, Steinfeld2009, Waugh2006}. Our findings of these low $\nicefrac{\Delta}{\Gamma}$ ratios reveal that the ventilation there is of a more advective nature than previously assumed. Additionally, the first reliable mean ages $\Gamma$ in this deeper region can be derived due to the unique half-life of \Ar{39} and increase with depth from 200 to 400 years (Figure \ref{fig:fig3} b). The deepest ventilation regime with $\nicefrac{\Delta}{\Gamma}\sim 0.9$ is found for water mainly composed of North Atlantic Deep Water (NADW) and Antarctic Bottom Water (AABW). We identify an increase in mean age up to $\Gamma = 800$ years at 3000 meters depth, followed by a decrease towards the ocean floor due to the better ventilated AABW. In these areas the TTD can only be poorly constrained due to low CFC-12 signals, resulting in large uncertainty of the inferred mean age $\Gamma = 810_{-320}^{+1200}$ years for our oldest sample at 3000 meters depth. Alternatively, using the extracted mean $\nicefrac{\Delta}{\Gamma}\sim 0.9$ for the third ventilation regime we find $\Gamma = 754_{-115}^{+138}$ years based on our \Ar{39} data. Thus \Ar{39} provides essential information on ventilation ages for old waters.

\section*{Broader implications and prospects}
Once the age distribution of a water body is known, one can derive concentrations of substances which have not been or which cannot be measured directly, as long as their input functions are known. An important factor in global climate change is the storage of anthropogenic carbon (C\textsubscript{ant}) by the ocean. In the studied area we find a C\textsubscript{ant} concentration within the AAIW/LSW range for profile \#44/\#55 at 1000 meters of \SI{\sim6.9}{\umol\per\kg}, which is in contrast to \SI{\sim3.8}{\umol\per\kg} using the commonly assumed $\nicefrac{\Delta}{\Gamma}$ unity ratio. For the water column between 1000 and 2000 meters in ventilation regime II, this difference adds up to \SI{\sim 3}{\mol\per\m\squared} which should be compared to the total column inventory of \SI{\sim25}{\mol\per\m\squared} \cite{Waugh2006}.
 
The presented work is an example how applied quantum technology developed in the context of fundamental research in atomic physics contributes to the advance of other fields, such as oceanography. The demonstrated ArTTA method for small sample sizes makes the unique time scale of \Ar{39} accessible for large scale ocean surveys. This can increase our knowledge of ocean dynamics significantly and, with that, support ocean and climate modelling. There are many other areas in environmental sciences, such as glaciology, limnology and groundwater research, where the novel analytical capabilities and much smaller sample sizes will enable new applications and lead to new insights.

\section*{Acknowledgments}
We thank the captain and the crew of RV METEOR for excellent support, B. Bogner for supporting the sampling at sea and our scientific and technical group, as well as several student assistants for their help with fieldwork and laboratory measurements. We thank S. Beyersdorfer for his contribution to the preparation of the samples, the complete team of the KIP workshop for machining the apparatus, V. Rädle for carefully reading the manuscript, and R. Erven for the graphic design of Figure 1. This research was supported by the Deutsche Forschungsgemeinschaft as part of the Sonderforschungsbereich 754 “Climate-Biogeochemistry Interactions in the Tropical Ocean” and two joint projects (OB 164/11-1, AE 93/14-1 and OB 164/12-1, AE 93/17-1) as well as by the European Research Commission Advanced Grant EntangleGen (project ID 694561). S.E. gratefully acknowledges support through the scholarship program of the Studienstiftung des deutschen Volkes.

\textbf{Author Contribution:} A.K. and T.T. conducted the sampling, S.E. performed the \Ar{39}-analyses with assistance from Z.F., L.R. and M.S. T.T. and T.S. analysed the Transit Time Distribution. W.A. and M.O. supervised the project. All authors worked on the manuscript. 

\textbf{Competing interests:} The authors declare no competing interests

\textbf{Additional information:}  
The complete data set from the Meteor 116 cruise including the \Ar{39}-data can be found at CCHDO:\\ https://cchdo.ucsd.edu/cruise/06M320150501

\textbf{Correspondence and requests} for materials should be addressed to S.E..

\appendix*

\section*{Methods}

\subsection*{Atom Trap Trace Analysis for argon with small sample sizes} 
The basic concept of our atomic beam apparatus for argon follows the established route \cite{Jiang2012, Jiang2011, Ritterbusch2014} with some modifications as mentioned below. It consists of a RF discharge source, which prepares the atoms in the metastable state necessary for laser cooling. The source is cooled with liquid nitrogen to reduce the initial velocity spread of the atomic beam. The divergent atomic beam, with a longitudinal mean velocity of \SI{270 }{\meter\per\second}, is collimated with transverse laser cooling in a tilted mirror setup and focused by a magneto-optical lens. Subsequently the atoms with a maximum velocity of \SI{600}{\meter\per\second} are longitudinally slowed down in a \num{1.8} meters long increasing field Zeeman slower. The final velocity of the atoms leaving the Zeeman slower is around \SI{70}{\meter\per\second} and is chosen above the capture velocity of the MOT. Thus the atomic beam is prevented to diverge rapidly at the end of the Zeeman slower. An additional laser frequency builds a second longitudinal slowing stage together with the rising slope of the magnetic field of the MOT. The atoms are refocused there by the MOT and losses due to the divergence of the atomic beam are reduced significantly. Finally the atoms are trapped inside the MOT and detected by their fluorescence with an avalanche-photodiode with a high time resolution and spatially resolved on a CCD camera. The detection threshold is set so that the rate of falsely counted atoms due to background noise is kept below 1 atom in 100 hours. The corresponding detection efficiency is \SI{94}{\percent} of all trapped \Ar{39} atoms.

An additional \SI{802}{\nano\m} laser deexcites isotope-selectively \SI{99}{\percent} of all metastable \Ar{40} atoms to the ground state, which are responsible for half of the background light on the single atom detection. The photons (\SI{843}{\nm}) emitted by the deexciting \Ar{40} atoms are detected by a photodiode which allows the monitoring of the atomic beam flux. Three differential pumping tubes together with seven turbo molecular pumps build up a pressure gradient from \SI{4.8e-6}{\milli\bar} in the source chamber down to \SI{1.4e-7}{\milli\bar} inside the MOT. Thereby, the argon throughput of the source is \SI{\sim 50}{mL\;STP} per hour.

By closing the vacuum completely (i.e. the argon sample recirculates in the system and no roughing pump removes gas from the system) the required sample size can be reduced by more than two orders of magnitudes. In this so called recycling or closed mode, the sample circulates permanently inside the system. Outgassing reactive gases, such as nitrogen and water, are removed by a getter pump, which does not affect noble gases. The argon sample will remain in the vacuum chamber and stays clean during the measurement process. A stable operation over the required measurement time is possible with samples sizes \SI{> 1}{mL\;STP} of argon. 

For further optimisation, the pressure inside the differential pumping tubes was simulated by a Monte Carlo simulation. By doubling the total pumping speed we achieve an atmospheric count rate of up to \SI{7.0}{atoms\per hour}, which is an improvement by a factor of 2 compared to the count rate reported previously\cite{Ritterbusch2014}.

\subsection*{Extraction and separation of argon}
The propane gas bottles containing the collected water samples were shipped to the laboratory in Heidelberg where each sample was degassed and purified. The dissolved gas was extracted by shaking the sample container and trapping the gas on a liquid nitrogen cooled activated charcoal trap. After 15 minutes more than \SI{95}{\percent} of the gas is stored on the trap. In a second step, all reactive gases are removed on a \SI{900}{\degreeCelsius} titanium sponge getter, while the released hydrogen is trapped on a second titanium getter at room temperature. The final gas fraction only consists of noble gases, thus of \SI{>98}{\percent} argon. With a getter capacity of about \SI{8}{\L\;STP} for the relevant reactive gases, more than 10 samples of \num{\sim 50} litres of water can be purified before having to replace the getter material.

Purifications of blank samples and helium leak tests are performed regularly and showing that cross sample contamination and leakage into the vacuum chamber is negligible. 

Two ocean samples were purified per day yielding between \num{5} to \SI{8}{mL\;STP} of argon per sample with a purity and extraction efficiency of \SI{>98}{\percent}.

\subsection*{\texorpdfstring{\Ar{39}-Analysis}{Ar-39-Analysis}}
The purified argon samples were analysed with the atom optical detection technique ArTTA optimised for \Ar{39} and small sample sizes as described above. Our 24 hour measurement cycle consists of 20 hours of measuring an ocean sample, followed by 2 hours of referencing with an enriched sample of a well-known concentration of $9.60_{-0.31}^{+0.33}$-times the atmospheric concentration. Finally the system is flushed for one hour by running the discharge source with krypton to avoid any significant cross sample contamination between the reference and the next environmental sample. After 24 hours the next sample is analysed. The currently achieved atmospheric \Ar{39} count rate of up to 7.0 atoms per hour enables – depending on the sample concentration – up to 150 counts for both the ocean sample and the reference within one day, leading to \SI{10}{\percent} uncertainty. For analysing the data, both reference measurements before and after counting the ocean sample are taken into account, resulting in an uncertainty of about \SI{7}{\percent} for the reference. Due to previous optimisation and characterisation of the apparatus with one million times enriched \Ar{39} samples more than five years ago there is still a low but detectable contamination present. This contamination together with a potential cross sample contamination was quantified with \Ar{39}-free underground samples originating from \ch{CO2} production wells\cite{Xu2015}. The measured count rates were corrected for this effect corresponding to about 10 atoms during a 20 hour measurement.

We apply a Bayesian approach for the analysis of the \Ar{39}-measurements. The reported concentrations are the most probable values and the bounds of the one sigma intervals contain most likely \SI{68.3}{\percent} of the values.

\subsection*{CFC measurements}
Two purge-and-trap gas chromatographic systems were used for the measurements of the transient tracer CFC-12 during the cruise, slightly modified from \cite{Bullister2008}. The traps for both systems consisted of \SI{100}{cm} \SI{1/16}{''} tubing packed with 70 cm Heysep D kept at temperatures between \num{-60} and \SI{-68}{\degreeCelsius} during trapping. The gas was desorbed by heating the trap to \SI{130}{\degreeCelsius} and was passed onto the pre-column consisting of \SI{20}{\cm} Porasil C followed by \SI{20}{\cm} Molsieve 5A in a \SI{1/8}{''} stainless steel column. A \SI{1/8}{''} packed main column consisting of \SI{180}{\cm} Carbograph 1AC (60-80 mesh) and a \SI{50}{\cm} Molsieve 5A post-column provided chromatographic separation. All columns were kept at \SI{50}{\degreeCelsius} and detection was performed on an Electron Capture Detector (ECD). 

The water samples for the determination of CFC-12 were collected in \SI{250}{\milli\L} ground glass syringes, of which an aliquot of about \SI{200}{\milli\L} was injected to the purge-and-trap system. Standardization was performed by injecting small volumes of gaseous standard with the CFC-12 calibrated to the SIO98 scale. The precision was determined to \SI{8}{\femto\mol\per\kg} and the limit of detection to \SI{0.06}{\femto\mol\per\kg}.  
Since we did not take samples for CFCs on the CTD casts where we collected samples for \Ar{39} measurements, we use nearby profiles that are comparable. Profile \#54 (N \SI {11.00}{\degree}, W \SI{22.00}{\degree}) was used to compare the CFC-12 values to the \Ar{39} profiles \#44 (N \SI{11.55}{\degree}, W \SI{23.00}{\degree}) and \#55 (N \SI{11.25}{\degree}, W \SI{22.00}{\degree}). For profile \#82 (N \SI{17.58}{\degree}, W \SI{24.30}{\degree}) we used a CTD taken immediately prior on the same position. Since the sampling for \Ar{39} and CFC-12 took place on different CTD-casts, we interpolated the CFC-12 data vs. density to match the \Ar{39} samples.


\begin{thebibliography}{10}

	\bibitem{IPCC}
	IPCC.
	\newblock {\em {Climate Change 2013: The Physical Science Basis. Contribution
	  of Working Group I to the Fifth Assessment Report of the Intergovernmental
	  Panel on Climate Change}}.
	\newblock Cambridge University Press, Cambridge, United Kingdom and New York,
	  NY, USA, 2013.
	
	\bibitem{Broecker1982}
	W.~Broecker, T.~Peng, and Z.~Beng{, and La\-mont-Do\-herty Geological
	  Observatory}.
	\newblock {\em Tracers in the sea}.
	\newblock La\-mont-Do\-herty Geological Observatory, Columbia University, 1982.
	
	\bibitem{Stoeven2015}
	T.~Stöven, T.~Tanhua, M.~Hoppema, and J.~L. Bullister.
	\newblock Perspectives of transient tracer applications and limiting cases.
	\newblock {\em Ocean Science}, 11(5):699--718, 2015.
	
	\bibitem{Fine2011}
	R.~A. Fine.
	\newblock {Observations of CFCs and {SF$_6$} as Ocean Tracers}.
	\newblock {\em Annual Review of Marine Science}, 3(1):173--195, 2011.
	
	\bibitem{Matsumoto2007}
	K.~Matsumoto.
	\newblock Radiocarbon-based circulation age of the world oceans.
	\newblock {\em Journal of Geophysical Research: Oceans}, 112(C9), 2007.
	
	\bibitem{Loosli1983}
	H.~H. Loosli.
	\newblock A dating method with $^{39}\text{Ar}$.
	\newblock {\em Earth and Planetary Science Letters}, 63(1):51--62, 1983.
	
	\bibitem{Loosli1968}
	H.~H. Loosli and H.~Oeschger.
	\newblock Detection of $^{39}\text{Ar}$ in atmospheric argon.
	\newblock {\em Earth and Planetary Science Letters}, 5:191--198, 1968.
	
	\bibitem{Broecker2000}
	W.~S. Broecker and T.~H. Peng.
	\newblock Comparison of $^{39}\text{Ar}$ and $^{14}\text{C}$ ages for waters in
	  the deep ocean.
	\newblock {\em Nuclear Instruments and Methods in Physics Research Section B:
	  Beam Interactions with Materials and Atoms}, 172(1):473--478, 2000.
	
	\bibitem{Holzer2010}
	M.~Holzer and F.~W. Primeau.
	\newblock {Improved constraints on transit time distributions from argon 39: A
	  maximum entropy approach}.
	\newblock {\em Journal of Geophysical Research: Oceans}, 115(C12), 2010.
	
	\bibitem{Schlosser1994}
	P.~Schlosser, B.~Kromer, R.~Weppernig, et~al.
	\newblock {The distribution of $^{14}\text{C}$ and $^{39}\text{Ar}$ in the
	  Weddell Sea}.
	\newblock {\em Journal of Geophysical Research: Oceans}, 99(C5):10275--10287,
	  1994.
	
	\bibitem{Schlitzer1985}
	R.~Schlitzer, W.~Roether, U.~Weidmann, P.~Kalt, and H.~H. Loosli.
	\newblock {A meridional $^{14}\text{C}$ and $^{39}\text{Ar}$ section in
	  northeast Atlantic deep water}.
	\newblock {\em Journal of Geophysical Research: Oceans}, 90(C4):6945--6952,
	  1985.
	
	\bibitem{Collon2004}
	P.~Collon, W.~Kutschera, and Z.-T. Lu.
	\newblock Tracing noble gas radionuclides in the environment.
	\newblock {\em Annual Review of Nuclear and Particle Science}, 54(1):39--67,
	  2004.
	
	\bibitem{Chen1999}
	C.~Y. Chen, Y.~M. Li, K.~Bailey, et~al.
	\newblock {Ultrasensitive Isotope Trace Analyses with a Magneto-Optical Trap}.
	\newblock {\em Science}, 286(5442):1139--1141, 1999.
	
	\bibitem{Jiang2012}
	W.~Jiang, K.~Bailey, Z.~T. Lu, et~al.
	\newblock An atom counter for measuring $^{81}\text{Kr}$ and $^{85}\text{Kr}$
	  in environmental samples.
	\newblock {\em Geochimica et Cosmochimica Acta}, 91:1--6, 2012.
	
	\bibitem{Sturchio2004}
	N.~C. Sturchio, X.~Du, R.~Purtschert, et~al.
	\newblock {One million year old groundwater in the Sahara revealed by
	  krypton-81 and chlorine-36}.
	\newblock {\em Geophysical Research Letters}, 31(5), 2004.
	
	\bibitem{Aggarwal2014}
	P.~K. Aggarwal, T.~Matsumoto, N.~C. Sturchio, et~al.
	\newblock Continental degassing of $^{4}\text{He}$ by surficial discharge of
	  deep groundwater.
	\newblock {\em Nature Geoscience}, 8:35, 2014.
	
	\bibitem{Gerber2017}
	C.~Gerber, R.~Vaikmäe, W.~Aeschbach, et~al.
	\newblock {Using $^{81}\text{Kr}$ and noble gases to characterize and date
	  groundwater and brines in the Baltic Artesian Basin on the one-million-year
	  timescale}.
	\newblock {\em Geochimica et Cosmochimica Acta}, 205:187--210, 2017.
	
	\bibitem{Buizert2014}
	C.~Buizert, D.~Baggenstos, W.~Jiang, et~al.
	\newblock {Radiometric $^{81}\text{Kr}$ dating identifies 120,000-year-old ice
	  at Taylor Glacier, Antarctica}.
	\newblock {\em Proceedings of the National Academy of Sciences}, 111(19):6876,
	  2014.
	
	\bibitem{Jiang2011}
	W.~Jiang, W.~Williams, K.~Bailey, et~al.
	\newblock {$^{39}\text{Ar}$ Detection at the $10^{-16}$ Isotopic Abundance
	  Level with Atom Trap Trace Analysis}.
	\newblock {\em Physical Review Letters}, 106(10):103001, 2011.
	
	\bibitem{Ritterbusch2014}
	F.~Ritterbusch, S.~Ebser, J.~Welte, et~al.
	\newblock {Groundwater dating with Atom Trap Trace Analysis of
	  $^{39}\text{Ar}$}.
	\newblock {\em Geophysical Research Letters}, 41(19):6758--6764, 2014.
	
	\bibitem{Lu2014}
	Z.~T. Lu, P.~Schlosser, W.~M. Smethie, et~al.
	\newblock Tracer applications of noble gas radionuclides in the geosciences.
	\newblock {\em Earth-Science Reviews}, 138:196--214, 2014.
	
	\bibitem{WebpageGeomar}
	http://cvoo.geomar.de/.
	
	\bibitem{Weiss1970}
	R.~F. Weiss.
	\newblock The solubility of nitrogen, oxygen and argon in water and seawater.
	\newblock {\em Deep Sea Research and Oceanographic Abstracts}, 17(4):721--735,
	  1970.
	
	\bibitem{Khatiwala2001}
	S.~Khatiwala, M.~Visbeck, and P.~Schlosser.
	\newblock {Age tracers in an ocean GCM}.
	\newblock {\em Deep Sea Research Part I: Oceanographic Research Papers},
	  48(6):1423--1441, 2001.
	
	\bibitem{Hall1994}
	T.~M. Hall and R.~A. Plumb.
	\newblock Age as a diagnostic of stratospheric transport.
	\newblock {\em Journal of Geophysical Research: Atmospheres},
	  99(D1):1059--1070, 1994.
	
	\bibitem{Waugh2003}
	D.~W. Waugh, T.~M. Hall, and T.~W.~N. Haine.
	\newblock Relationships among tracer ages.
	\newblock {\em Journal of Geophysical Research: Oceans}, 108(C5), 2003.
	
	\bibitem{Stoeven2014}
	T.~Stöven and T.~Tanhua.
	\newblock {Ventilation of the Mediterranean Sea constrained by multiple
	  transient tracer measurements}.
	\newblock {\em Ocean Science}, 10(3):439--457, 2014.
	
	\bibitem{Bullister2017}
	J.~L. Bullister.
	\newblock {Atmospheric Histories (1765-2015) for {CFC-11, CFC-12, CFC-113,
	  CCl$_4$, SF$_6$ and N$_2$O} (NCEI Accession 0164584)}.
	\newblock {\em NOAA National Centers for Environmental Information}, 2017.
	
	\bibitem{Schneider2012}
	A.~Schneider, T.~Tanhua, A.~Körtzinger, and D.~W.~R. Wallace.
	\newblock {An evaluation of tracer fields and anthropogenic carbon in the
	  equatorial and the tropical North Atlantic}.
	\newblock {\em Deep Sea Research Part I: Oceanographic Research Papers},
	  67:85--97, 2012.
	
	\bibitem{Steinfeld2009}
	R.~Steinfeldt, M.~Rhein, J.~L. Bullister, and T.~Tanhua.
	\newblock {Inventory changes in anthropogenic carbon from 1997–2003 in the
	  Atlantic Ocean between 20°S and 65°N}.
	\newblock {\em Global Biogeochemical Cycles}, 23(3), 2009.
	
	\bibitem{Waugh2006}
	D.~W. Waugh, T.~M. Hall, B.~I. McNeil, R.~Key, and R.~J. Matear.
	\newblock Anthropogenic {CO$_2$} in the oceans estimated using transit time
	  distributions.
	\newblock {\em Tellus B: Chemical and Physical Meteorology}, 58(5):376--389,
	  2006.
	
	\bibitem{Xu2015}
	J.~Xu, F.~Calaprice, C.~Galbiati, et~al.
	\newblock A study of the trace $^{39}\text{Ar}$ content in argon from deep
	  underground sources.
	\newblock {\em Astroparticle Physics}, 66:53--60, 2015.
	
	\bibitem{Bullister2008}
	J.~L. Bullister and D.~P. Wisegarver.
	\newblock The shipboard analysis of trace levels of sulfur hexafluoride,
	  chloro\-fluoro\-carbon-11 and chloro\-fluoro\-carbon-12 in seawater.
	\newblock {\em Deep Sea Research Part I: Oceanographic Research Papers},
	  55(8):1063--1074, 2008.

\end{thebibliography}

\end{document}